\documentclass[12pt]{book}

\usepackage[dvips]{graphicx,color}
\usepackage{makeidx,tsukuba}
\usepackage{times}
\makeauthorindex
\makeindex

\begin{document}

\BookTitle{\itshape The 28th International Cosmic Ray Conference}
\CopyRight{\copyright 2003 by Universal Academy Press, Inc.}
%\tableofcontents
\pagenumbering{arabic}

%%%%%%%%%%% The first letter of each word should be capital letter.
\chapter{Very High Energy Observations Of PSR B1823-13}

\author{%
%
% You can include as many co-authors as you wish, unless
% the title/author information fits within 1 page.
%
T.A.~Hall,$^{1,2}$ I.H.~Bond, P.J.~Boyle, S.M.~Bradbury, J.H.~Buckley,
D.~Carter-Lewis, O.~Celik, W.~Cui, M.~Daniel, M.~D'Vali,
I.de~la~Calle~Perez, C.~Duke, A.~Falcone, D.J.~Fegan, S.J.~Fegan,
J.P.~Finley, L.F.~Fortson, J.~Gaidos, S.~Gammell, K.~Gibbs,
G.H.~Gillanders, J.~Grube, J.~Hall, D.~Hanna, A.M.~Hillas,
J.~Holder, D.~Horan, A.~Jarvis, M.~Jordan, G.E.~Kenny, M.~Kertzman,
D.~Kieda, J.~Kildea, J.~Knapp, K.~Kosack, H.~Krawczynski, F.~Krennrich,
M.J.~Lang, S.~LeBohec, E.~Linton, J.~Lloyd-Evans, A.~Milovanovic,
P.~Moriarty, D.~Muller, T.~Nagai, S.~Nolan, R.A.~Ong, R.~Pallassini,
D.~Petry, B.~Power-Mooney, J.~Quinn, M.~Quinn, K.~Ragan, P.~Rebillot,
P.T.~Reynolds, H.J.~Rose, M.~Schroedter, G.~Sembroski, S.P.~Swordy,
A.~Syson, V.V.~Vassiliev, S.P.~Wakely, G.~Walker, T.C.~Weekes,
J.~Zweerink \\
{\it
(1) University of Arkansas at Little Rock, Dept. of Physics and Astronomy, Little Rock, AR 72204, USA
(2) The VERITAS Collaboration--see S.P.Wakely's paper} ``The VERITAS
Prototype'' {\it from these proceedings for affiliations}
}%% end of author

\section*{Abstract}
To date three plerionic systems have been detected as emitters of very
energetic photons.  As part of an ongoing study of pulsar systems at
the Whipple observatory, observations of the plerion PSR B1823-13 are
being conducted. Observations were made with the Whipple 10~m
gamma-ray telescope utilizing the high resolution, 490 pixel camera.

\section{Introduction}

Four sources of detected very high energy (VHE) gamma radiation are
associated with systems containing pulsars.  Three of these sources
(Crab Nebula [11], Vela pulsar [12], and PSR B1706-44 [6]) represent a
class of objects known as plerions.  A plerion is a compact nebula
resulting from a relativistic particle wind emanating from a pulsar
interacting with the pulsar's environment.

PSR B1823-13 is a young, spin-powered pulsar with characteristics
similar to both Vela and PSR B1706-44. Observations in the X-ray band [3]
have revealed that PSR B1823-13 is powering a compact synchrotron
nebula (plerion) with a physical size similar to that seen in the
other plerionic sources (Crab, Vela, and PSR B1706-44). PSR B1823-13
is 7th in a rank ordered list of $\dot{E}$/d$^2$.

In addition to the X-ray detection of this young isolated neutron
star, the position of PSR B1823-13 is near the EGRET unidentified
source, 3EG 1823-1314 [5].  This EGRET source is also one of the stronger
detections in the Lamb 1 GeV catalog [7].  The above factors make PSR
B1823-13 a prime candidate for observations in the VHE band.

% % % % % % % % % % % % % % % % % % % % % % % % % % % % % % % % % % % % 
% 
% All papers should be written in English.  The title should be followed
% by the name(s) of the author(s), the institution and its address as an
% above example.  The outline of the paper should be set out using
% appropriate headings, i.e.\ Abstract, 1.\ Introduction, 2.\ Methods, 3.\
% Results, 4.\ Discussion, and 5.\ Conclusions.  Your manuscript should
% include all the files listed below and these files must be archived to a
% 0
% single file for submission through ICRC2003 web (individual) page: \\
% http://www-rccn.icrr.u-tokyo.ac.jp/icrc2003/proceedings.html \\ 
% ``How to prepare your manuscript" for submission is described in detail
% there. In order to quickly publish the proceedings of papers, the
% manuscript should be submitted on and before the deadline: \fbox{16 May,
% 2003}.

% \begin{itemize}
% \item Electronic version of the manuscript
%    \begin{enumerate}
%       \item  \LaTeX\ file
%       \item  EPS (or PS) files of figures
%       \item  PDF file (Final layout of your contribution containing the text
% and all figures)
%    \end{enumerate}
% \end{itemize}
% 

\section{Observations and Analysis}
The observations of PSR B1823-13 presented here were made with the
Whipple 10~m gamma-ray telescope located on Mt. Hopkins in southern
Arizona [2].  These data were collected during the spring of
2000. During this time, the telescope utilized a high resolution
camera consisting of an array of 490 photomultiplier tubes mounted at
the focal plane of the reflector. A detailed description of the
telescope and its characteristics can be found in [2, 4].

Images of extensive air showers initiated by high-energy photons and
cosmic rays are made by recording the \v{C}erenkov radiation emitted as
the shower propagates through the atmosphere.  By making use of
distinctive differences in the angular distribution of light and the
orientation of the shower images, it is possible to differentiate a
gamma ray initiated event from a very large hadronic background.

\subsection{Unpulsed Analysis}

The observations reported here were collected in the ON/OFF
observation mode.  In ON/OFF mode, the candidate source position is
observed for 28 minutes (ON run) followed by a 28 minute reference
observation (OFF run) taken at the same azimuth and elevation as the
ON run.  The OFF region is used to estimate the background counts in
the ON region.

% TRACKING data are
% collected by observing the putative source position, with the
% background estimated from events whose arrival direction is not
% consistent with the source position [1]. 

During the 2000 observing season, 430 minutes of ON source data (16
on/off pairs) were acquired.  
% Due to the low elevation of PSR B1823-13
% (less than 45 degrees), bright source data (Crab Nebula and Markarian
% 421) at similar elevations were used to optimize cuts suited to the
% elevations of PSR B1823-13. Table 1 summarizes the values of the cuts
% used.  
A moment analysis routine [9] utilizing the parameter cuts was applied
to the data yielding a 3.4 $\sigma$ excess (0.7 g/min) in an ALPHA
analysis.  Figure 1 displays the ALPHA distribution and the ON minus
OFF counts as a function of ALPHA angle.  A flux upper limit of
2.6x10$^{-11}$cm$^{-2}$s$^{-1}$ at a peak energy of 1 TeV was derived
at a coinfidence level of 99.9\%.  A 2-Dimensional analysis [8] was
also applied to the data yielding a 3.1 sigma excess near the position
of PSR B1823-13 as seen in Figure 2.

% \begin{table}[t]
% \caption{Optimized parameter cuts for low elevation sources.}
% \begin{center}
%\begin{tabular}{l|ccccc}
%\hline
%Parameter & Lower Bound & Upper Bound \\
%\hline
%Max1        & 30 d.c.      & N/A    \\
%Max2        & 30 d.c.      & N/A    \\
%Length      & 0.13$^\circ$ & 0.25$^\circ$    \\
%Width       & 0.05$^\circ$ & 0.12$^\circ$    \\
%Distance    & 0.4$^\circ$  & 1.0$^\circ$    \\
%Length/Size & N/A          & 0.00045    \\
%Alpha       & 0$^\circ$    & 15$^\circ$    \\

% \hline
% \end{tabular}
% \end{center}
% \end{table}

% However, standard analysis methods are not optimized for low elevation
% sources.  Due to the low elevation of PSR B1823-13 (less than 45
% degrees), Crab Nebula data at similar elevations were used to optimize
% cuts better suited to the elevations of PSR B1823-13. Table 1
% summarizes the values of the cuts used. Applying these new cuts to the
% ON/OFF data yields a 3.7 sigma excess (0.8 g/min) in an alpha analysis
% (see Figure 1) and a 3.5 sigma excess in a 2-D analysis as seen in
% Figure 2.

 \begin{figure}[!h]
   \begin{center}
    \includegraphics[height=13pc]{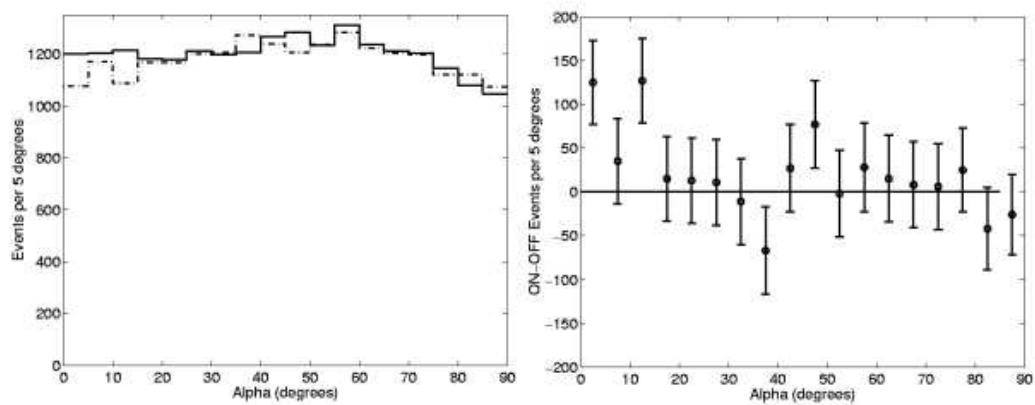}
   \end{center}
   \vspace{-0.5pc}
   \caption{Results of the ALPHA analysis of PSR B1823-13.  Plots show the ALPHA distribution and the ON minus OFF counts in 5$^{\circ}$ bins.  }
 \end{figure}

\begin{figure}[h!]
   \begin{center}
    \includegraphics[height=14.5pc]{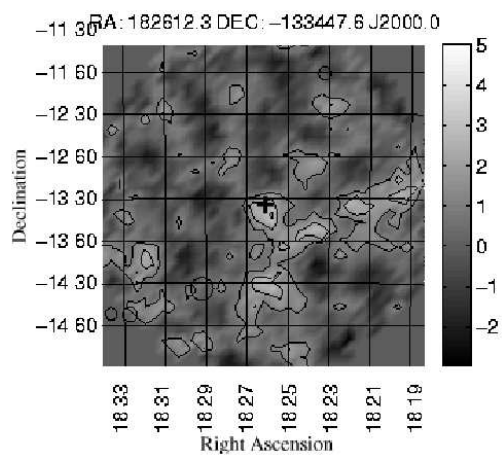}
   \end{center}
   \vspace{-0.5pc}
   \caption{2D analysis of PSR B1823-13.  Curves represent 1$\sigma$ contours. The position of PSR B1823-13 is indicated with a cross.}
 \end{figure}

%  \begin{figure}[t]
 %   \begin{center}
%     \includegraphics[height=13.5pc]{psrfig1_2d.eps}
%    \end{center}
%    \vspace{-0.5pc}
%    \caption{The asymmetry $A_2$ as a function of $x$.}
%  \end{figure}

% Approximately 25 TRACKING observations were made of PSR B1823-13
% (these include ON observations).  These were analyzed using a tracking
% ratio \cite{catXXX}of 0.304+/- 0.0062, obtained from the OFF source
% data sets of PSR B1823-13.  Applying the cuts optimized for low
% elevations to the data gives a 5.7 $\sigma$ (0.74 g/min).  However, a
% smaller result (~XXX $\sigma$) is obtained when errors in the tracking
% ratio (which are very large due to the limited sample size) are
% incorporated.

\subsection{Periodic Analysis}
\label{periodic}

The arrival times of \v{C}erenkov events were recorded by a GPS clock and
a 10 MHz oscillator calibrated by a GPS second mark to achieve an
absolute time resolution of 0.1 $\mu$s.  All arrival times were then
transformed to the solar system barycenter using the JPL DE200
planetary ephemerides [10].  The phase of each event passing the
parameter cuts was calculated using the appropriate ephemeris for the
epoch of the data.  To test for the presence of a periodic signal,
$\chi^2$ and $Z^2_m$ tests were performed. The analysis of event
arrival times showed no evidence of modulation at the pulsar period.

In order to calculate an upper limit for pulsed emission, a method
utilizing the $Z^2_2$ statistic, which assumes a sinusoidal pulse
profile, was used.  The pulsed fraction at a peak
energy of 1 TeV is found to be less than 6\%
of the background at a confidence level of 99.9\%. 
% Table 3 shows the results of the periodic analysis
% performed on the pulsar data.

% All paper should be assigned 4 pages.%%%%

% % % % % % % % % % % % % % % % % % % % % % % % % % % % % % % % % % % % 
\section{Discussion}

Results from the observations of PSR B1823-13 show an excess in the
ALPHA distribution consistent with a weak source. Figure 3 shows the
cumulative significance obtained from the analysis.  The gradual
increase in significance may be due to a weak, persistent gamma-ray
source.
%  If this excess is due to a gamma-ray signal, it would imply a
%flux of 1.4x10$^{-11}$cm$^{-2}$s$^{-1}$ at a peak energy of 1 TeV.
Further observations are currently being conducted.  The results of
these new observations will be presented at the conference.

\begin{figure}[t]
   \begin{center}
    \includegraphics[height=13.5pc]{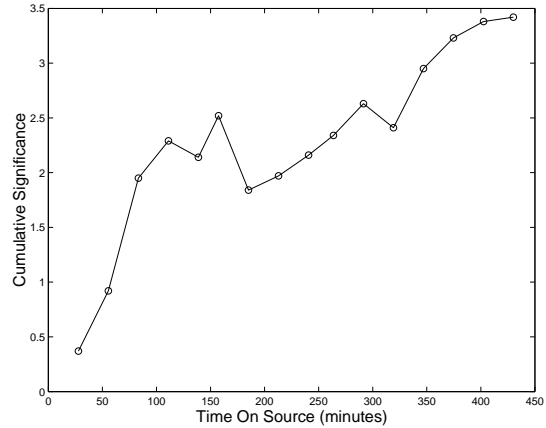}
   \end{center}
   \vspace{-0.5pc}
   \caption{Cumulative significance of the observations of PSR B1823-13.}
 \end{figure}

\section{Acknowledgments}
We acknowledge the technical assistance of E. Roache and J. Melnick.  We would also like to thank A. Lyne and Z. Arzoumanian for providing radio ephemerides for this report.  This research is supported by grants from the U. S. Department of Energy, by Enterprise Ireland, and by PPARC in the UK.

\section{References}

% A list of references should be placed at the end of the paper.  Use
% SINGLE LINE spacing and start each reference on a new line.  References
% are to be listed alphabetically by the last name of the first author.
% When two or more references by the same author are listed, the year of
% publication is be taken into account with the earliest work being listed
% first. See the below example.  The citation to the references can be
% written as [3] or [2, 4].

\vspace{\baselineskip}

\re
1.\ Catanese M. \ et al.\ 1998, ApJ 501, 616
\re
2.\ Cawley M.F., Fegan D.J., Harris K., Kwok P.W., Hillas A.M. \ 1990, Experimental Astronomy 1, 173
\re
3.\ Finley J.P., Srinivasan R., Park S. \ 1996, ApJ 466, 938
\re
4.\ Finley J.P. \ et al.\ 2001, in Proc. 27th ICRC (Hamburg, Germany), OG 2.5
\re
5.\ Hartman R.C., \ et al.\ 1999, ApJS 123, 79
\re
6.\ Kifune T. \ et al.\ 1995, ApJ 438, L91
\re
7.\ Lamb R.C., Macomb D.J. \ 1997, ApJ 488, 872
\re
8.\ Lessard R.W., Buckley J.H., Connaughton V., Le Bohec, S. \ 2001, APh 15, 1
\re
9.\ Reynolds P.T. \ et al.\ 1993, ApJ 404, 206
\re
10.\ Standish E.M. Jr.\ 1982, AAP, 114, 297
\re
11.\ Weekes T.C.\ et al.\ 1989, ApJ 342, 379
\re
12.\ Yoshikoshi T.\ et al.\ 1997, ApJ 487, L65

% \re
% 1.\ Ikegami H.\ 1997, MNRAS 287, 651
% \re
% 2.\ Ikegami H., Shibata T.\ 1985, A\&Ap 39, 941
% \re
% 3.\ Ikegami H., Shibata T., Tanaka Y., White N.E.\ 1991, ApJ 320, L127
% \re
% 4.\ Ikegami H.\ et al.\ 1987, in Black Hole, ed.\ Wheeler (Universal Academy
% Press, Tokyo)

% \section{How to Get Style-File for {\LaTeX} and Sample File}

% You can get the Instructions for the Preparation of
% Manuscripts for the Symposium Proceedings
% through ICRC2003 web (individual) page: \\
% http://www-rccn.icrr.u-tokyo.ac.jp/icrc2003/proceedings.html.
% \begin{itemize}
% \item File Name:
% 	\begin{itemize}
% 	\item icrc.pdf (instruction in PDF format)
% 	\item icrc.tex (instruction by \LaTeX\ source)
% 	\item tsukuba.sty (style file for \LaTeX)
% 	\item figure.eps (figure sample)
% 	\end{itemize}
% \end{itemize}

% \section{Attention}

% \begin{itemize}
% \item
% Please do not use {\boldmath $\backslash$}{\bf newcommand} and {\boldmath
% $\backslash$}{\bf def} on the \LaTeX\ file.

% \item
% If you have any question for the proceedings,
 % please do not hesitate to contact us.
% \end{itemize}

% For inquiry about the style file,
% contact to :
% \begin{center}
% Universal Academy Press, Inc.\\
% http://www.uap.co.jp \\
% {\bf e-mail: icrc@uap.co.jp}
% \end{center}

% For general inquiry on proceedings,
% contact to :
% \begin{center}
% ICRC 2003 Local Organizing Committee proceedings desk\\
% {\bf e-mail: icrcproc@icrr.u-tokyo.ac.jp}
% \end{center}

\endofpaper
\end{document}